\documentclass[aip,pop,twocolumn, 10pt, floatfix]{revtex4-1} 

\usepackage{amsmath}
\usepackage{amssymb}
\usepackage{epsfig}
\usepackage{bm}

\begin{document} 
\title{ Simulating Gyrokinetic Microinstabilities in Stellarator Geometry with GS2 } 

\renewcommand{\thefootnote}{\alph{footnote}}

\author{J. A. Baumgaertel}%
\affiliation{%
 Princeton Plasma Physics Laboratory, Princeton, New Jersey 08543
}%
\author{E. A. Belli }
\affiliation{%
 General Atomics, San Diego, California 92186 
}%
\author{W. Dorland}
\affiliation{%
University of Maryland, College Park, Maryland 20742
}%
\author{W. Guttenfelder}
\affiliation{%
 Princeton Plasma Physics Laboratory, Princeton, New Jersey 08543
}%
\author{G. W. Hammett}
\affiliation{%
 Princeton Plasma Physics Laboratory, Princeton, New Jersey 08543
}%
\author{D. R. Mikkelsen}
\affiliation{%
 Princeton Plasma Physics Laboratory, Princeton, New Jersey 08543
}%
\author{G. Rewoldt}
\affiliation{%
 Princeton Plasma Physics Laboratory, Princeton, New Jersey 08543
}%
\author{W. M. Tang}
\affiliation{%
 Princeton Plasma Physics Laboratory, Princeton, New Jersey 08543
}%
\author{P. Xanthopoulos}
\affiliation{%
 Max-Planck-Institut fuer Plasmaphysik, D-17491 Greifswald, Germany 
}%

\date{\today}

\begin{abstract} 
The nonlinear gyrokinetic code GS2 has been extended to treat non-axisymmetric
stellarator geometry. Electromagnetic perturbations and multiple trapped particle regions
are allowed. Here, linear, collisionless, electrostatic simulations of the
quasi-axisymmetric, three-field period National Compact Stellarator Experiment (NCSX) design QAS3-C82
have been successfully benchmarked against the eigenvalue code FULL. Quantitatively, the linear stability calculations
of GS2 and FULL agree to within $\sim10\%$. 
\end{abstract}
\maketitle 

\section{Introduction}

One of the most important issues for magnetic fusion is the confinement
of heat and particles. Turbulent transport (most likely the
result of drift wave instabilities) causes a significant amount of
heat loss in tokamaks and spherical tori.\cite{liewer_measurements_1985}
Neoclassical transport, on the other hand, can often account for the poor
confinement in traditional stellarators.\cite{fu_ideal_2007} However,
modern stellarator designs, such as Wendelstein 7-AS (W7-AS),\cite{sapper_stellarator_1990}
Wendelstein 7-X (W7-X),\cite{beidler_physics_1990,grieger_physics_1992} the National Compact Stellarator Experiment (NCSX),\cite{zarnstorff_physics_2001} the Large Helical Device (LHD) \cite{yamada_configuration_2001}, and the Helically Symmetric Experiment (HSX) \cite{gerhardt_experimental_2005,canik_experimental_2007,talmadge_experimental_2008} have shown or
are designed to have improved neoclassical confinement and stability
properties. Understanding plasma turbulence and transport could further
improve the performance of stellarators. Progress in design of stellarators
for optimal transport has been made by coupling the gyrokinetic code
GENE\cite{jenko_electron_2000} with the configuration optimization code STELLOPT.\cite{mynick_optimizing_2010,reiman_physics_1999}

Gyrokinetic studies of drift-wave-driven turbulence in stellarator
geometry are relatively recent and comprehensive scans are scarce.
Most of these studies were done using upgraded versions of well-established
axisymmetric codes which include comprehensive kinetic dynamics (multispecies,
collisions, finite beta) to the more general case of non-axisymmetric
stellarator geometry, in the flux tube limit. The first non-axisymmetric
linear gyrokinetic stability studies, for both the ion-temperature-gradient-driven (ITG) mode and the trapped-electron mode (TEM),
were done with the linear eigenvalue FULL code,\cite{rewoldt_electromagnetic_1982,rewoldt_collisional_1987,rewoldt_drift_1999}
including a comparison of stability in nine stellarator configurations.\cite{rewoldt_comparison_2005}
Extensive studies have been done with the upgraded GENE code, including
the first nonlinear gyrokinetic simulations.\cite{xanthopoulos_nonlinear_2007} More recently, the GKV-X code, which uses the adiabatic electron approximation,
has been used to analyze linear ITG modes and zonal flows in LHD and nonlinear studies are in progress.\cite{watanabe_gyrokinetic_2007}

For this purpose, the axisymmetric nonlinear microinstability code
GS2\cite{dorland_electron_2000} has been extended to treat the more
general case of non-axisymmetric stellarator geometry. GS2
contains a full (except that the equilibrium distribution function is taken to be a Maxwellian) implementation of the 5-D Frieman
and Chen nonlinear gyrokinetic equation in the flux tube limit,\cite{dorland_electron_2000,howes_astrophysical_2006} with an efficient parallelization for modern supercomputers.\cite{kotschenreuther_comparison_1995}
It treats electrons and an arbitrary number of ion species on an equal
footing, and includes trapped particles, electromagnetic perturbations,
and a momentum-conserving pitch-angle-scattering collision operator.
The extension of the code to non-axisymmetric geometry not only retains
all of the above dynamics of the axisymmetric version, but also allows,
most importantly, multiple trapped particle regions and multiple
totally-trapped pitch angles at a given theta grid point. (By ``totally-trapped,'' we mean particles with such a small parallel velocity that they are limited to one grid point at the bottom of a well.) Tokamaks only have one trapped
particle region, but as stellarators can have many deep, narrow magnetic
wells which can trap particles (though NCSX has only a single deep
well, with other shallow wells, and is a bridge in configuration space
between tokamaks and other stellarators). In order to treat the trapped
particles accurately, one needs to resolve these wells sufficiently
with high grid resolution. With the GS2 modifications, we allow
for more flexible, decoupled pitch angle and parallel spatial grids, relative to the original GS2 algorithm which required every grid point ($\theta_j$) along the field line to correspond exactly to the turning point of trapped pitch angle ($\lambda_i=\mu/E$) grid points.\cite{kotschenreuther_comparison_1995}

Beyond these extensions, a GS2 stellarator simulation requires different
geometry codes to build its input grids than standard tokamak runs.
For these non-axisymmetric simulations, the geometrical coefficients
are based on a VMEC\cite{hirshman_improved_1990,hirshman_momcon:_1986}
3D MHD equilibrium, which is transformed into Boozer coordinates\cite{boozer_guiding_1980}
by the TERPSICHORE code.\cite{anderson_methods_1990} From this equilibrium,
the VVBAL code\cite{cooper_variational_1992} constructs data along
a chosen field line necessary for the microinstability calculations:
$B=|\mathbf{B}|$, the $\nabla \mathbf{B}$ drift, the curvature drift, and the metric
coefficients. While these extensions were used to study HSX plasmas,\cite{guttenfelder_effect_2008} here we verify
the non-axisymmetric extension of GS2 through comparisons with FULL on NCSX plasmas.
Good agreement between the GS2 code and the FULL code in the axisymmetric
limit has been extensively demonstrated previously.\cite{kotschenreuther_comparison_1995,bourdelle_stabilizing_2003}
While the non-axisymmetric upgrade of GS2 retains the nonlinear dynamics,
in these studies we focus on systematic scans of gyrokinetic linear
stability. 

The organization of this paper is as follows. The NCSX equilibrium
used for the benchmark is described in Section II. Comparisons between the GS2 code and the FULL code in non-axisymmetric geometry
over a range of parameters including  $\eta=L_{n}/L_{T}$ (where $L_{n}$ is the density gradient scale length and $L_{T}$
is the temperature gradient scale length), $k_{y}\rho_{i}$, $T_{i}/T_{e}$, and geometrical coordinates are presented in Section III. Further results using the GS2 code to investigate effects
of density and temperature gradients are presented in Section
IV. Conclusions and a discussion of future work are given in Section
V. Finally, Appendix A contains definitions of the normalizations and radial coordinate
used by GS2.

\section{The QAS3-C82 Equilibrium}

All of the benchmark calculations use a VMEC equilibrium based on
a 1999 NCSX design known as QAS3-C82,\cite{reiman_physics_1999}
which is shown in Figure 1. This configuration is quasi-axisymmetric
with three field periods. It has an aspect ratio of 3.5 and a major
radius of 1.4 m. NCSX was designed to have good neoclassical transport
and MHD stability properties and good drift trajectories
similar to those in tokamaks. Strong axisymmetric components of shaping
provide good ballooning stability properties at lower aspect ratio.
Furthermore, the QAS3-C82 equilibrium has a monotonically increasing
rotational transform profile which provides stability to neoclassical
tearing modes across the entire cross section.\cite{reiman_recent_2001,reiman_physics_1999}

\begin{figure}
\includegraphics[scale=1.0]{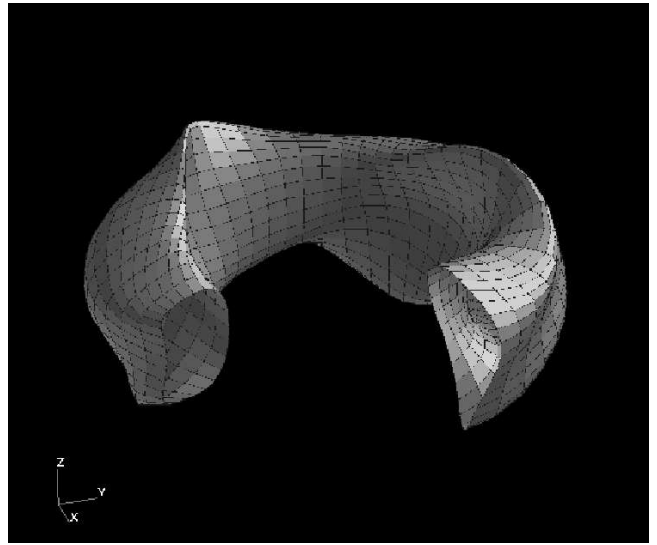}
\caption{Equilibrium of NCSX design QAS3-C82 which is quasi-axisymmetric and
has 3 field periods.}
\end{figure}

For most of these runs, we chose the surface at $s=0.875$
($s\sim\langle(r/a)^{2}\rangle$ is the normalized toroidal flux)
and the field line at $\alpha=\pi/3$ ($\alpha=\zeta-q\theta$; $\zeta$
is the Boozer toroidal angle, $\theta$ is the Boozer poloidal angle). The cross-section at this point is the crescent shape, seen in Figure 17 of Ref. \onlinecite{ku_modular_2010}. The coordinate
along the field line is $\theta$, the poloidal angle. At this surface,
the safety factor $q=2.118$ and the average $\beta$ (the
ratio of the plasma pressure to the magnetic pressure) is $\langle\beta\rangle=0.01\%$. Lastly, the ballooning parameter\cite{cooper_variational_1992} is $\theta_{0}=0$, except in Figure 6.

Figure 2 shows the variation of the magnitude of the magnetic field
along a chosen magnetic field line. Resolution studies for the spatial grid used in the GS2 runs indicate that 330
theta grid points per poloidal period and about
90 pitch angles ($\lambda=\mu/E$) showed convergence in the growth rate to within $2\%$, however $<10\%$ error is possible with coarser grids.
It was also found that a $\theta$ range extending from $-3\pi$ to
$3\pi$ was sufficient for a typical simulation grid, meaning that
the eigenfunctions for the modes decayed to insignificant values before
reaching these boundaries. (The endpoints of $B(\theta)$ were increased slightly, by less than $1\%$, to be global maxima, per normal GS2 operations.)

\begin{figure}
\includegraphics[scale=1.0]{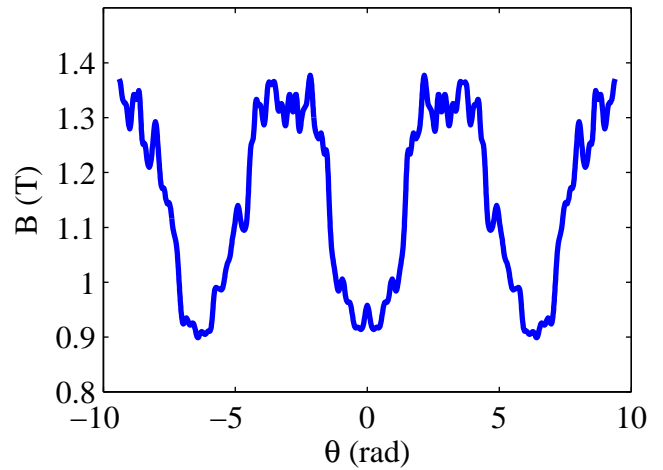}\caption{Standard B vs. $\theta$ grid for QAS3-C82, with $s=0.875$, $\alpha=\pi/3$,
and $\theta_{0}=0$.}

\end{figure}

The equilibrium's geometry suggests unstable drift waves exist. The
variations of $(k_{\perp}/n)^{2}$, where $n$ is the toroidal
mode number, and the curvature drift along the same chosen field line
can be seen in Figures 3 and 4. By convention, positive curvature
drifts are ``bad'' or destabilizing, while
negative curvature drifts are ``good'' or
stabilizing. Significant unstable modes occur where $k_{\perp}$
is small, which is near $\theta=0$ for this equilibrium, since instabilities
are generally suppressed at large $k_{\perp}$ by FLR averaging.
Also, because Figure 4 indicates that the curvature is bad in this region
near $\theta=0$, it is expected that unstable modes will appear
here.

\begin{figure}
\includegraphics[scale=1.0]{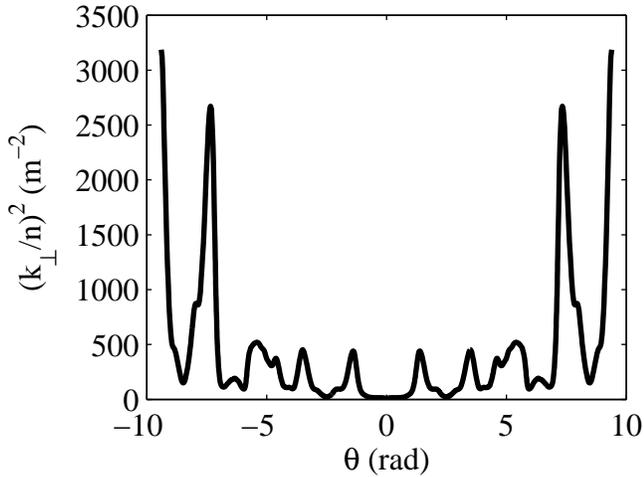}\caption{Variation of $(\frac{k_{\perp}}{n})^{2}(\theta)$ for QAS3-C82, with
$s=0.875$, $\alpha=\pi/3$, and $\theta_{0}=0$.}

\end{figure}

\begin{figure}

\centering
\includegraphics[scale=1.0]{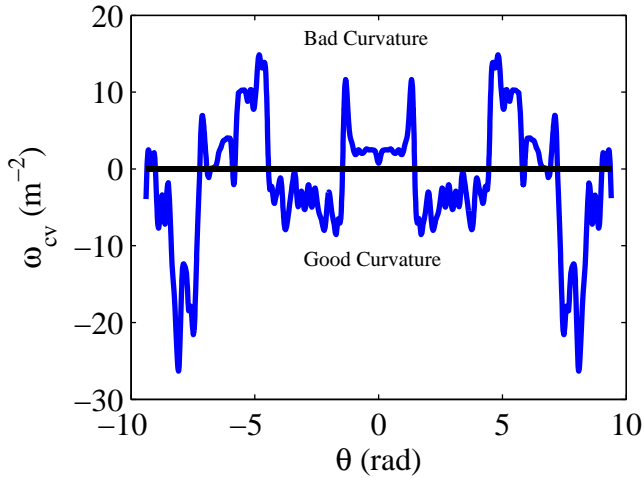}\caption{Variation of the curvature drift frequency ($\omega_{cv}=(\mathbf{k_\perp}/n)\cdot \mathbf{b}\times [\mathbf{b} \cdot \nabla \mathbf{b}]$) (for $n=1$) along $\theta$ for QAS3-C82, with
$s=0.875$, $\alpha=\pi/3$, and $\theta_{0}=0$. }

\end{figure}

\section{Benchmarks with FULL}

Comparisons between the GS2 code and the FULL code in non-axisymmetric
geometry over a range of parameters using the QAS3-C82 equilibrium
show linear agreement for our standard case, whose local parameters
are shown in Table I. The product of the perpendicular wave
number and the gyroradius at $\theta=0$, $k_{y}\rho_{i}$,
is $0.3983$ (where the toroidal mode number $n=25$; see App. A) for all cases
unless otherwise specified. The standard case is relatively close
to the edge, which accounts for the low values of ion temperature,
$T_{i}$, electron temperature, $T_{e}$, and relatively large values for the gradients. The parameter, $\eta=L_{n}/L_{T}$ is usually $\eta=3$, placing
most of our studies in an ITG regime (see Figure 7).
Correspondingly, $a_{N}/L_{ni}=a_{N}/L_{ne}=13.096$ and $a_{N}/L_{Ti}=a_{N}/L_{Te}=39.288$.
The major radius is approximately $R\approx1.4m$. The normalizing
scale length is $a_{N}=n/k_{\perp}(\theta=0)=0.352m$,
not the minor radius, and is described in detail in App. A. These studies are done with electrons and deuterium ions.

\begin{table}
\begin{tabular}{|c|c|}
\hline 
$s \approx \left(\langle r/a \rangle \right)^2$ & $0.875$\tabularnewline
\hline 
$\alpha=\zeta-q\theta$  & $\pi/3$\tabularnewline
\hline 
$\theta_{0}$  & $0$\tabularnewline
\hline 
$q$ & $2.118$\tabularnewline
\hline 
$\langle\beta\rangle$ & $0.01\%$\tabularnewline
\hline 
$k_{y}\rho_{i}$ & $0.3983$$(n=25)$\tabularnewline
\hline 
$T_{i}=T_{e}$ & $1keV$\tabularnewline
\hline 
$\eta_{i}=\eta_{e}$ & $3$\tabularnewline
\hline 
$a_{N}/L_{ni}=a_{N}/L_{ne}$ & $\approx13.096$\tabularnewline
\hline 
$a_{N}/L_{Ti}=a_{N}/L_{Te}$ & $\approx39.288$\tabularnewline
\hline 
$R$ & $\approx4a_{N}\approx1.4m$\tabularnewline
\hline 
$a_{N}=(\frac{n}{k_{\perp}(\theta=0,\theta_0=0)})$ & $\approx0.352m$\tabularnewline
\hline
$B_{a}=\langle B \rangle$ & $1.15 T$ \tabularnewline
\hline
$m_{ref}$ & $2m_p$\tabularnewline
\hline
$v_t$ & $\sqrt{(eT_i 1000)/m_{ref}}$\tabularnewline
\hline
GS2 $\omega$ units $v_t/a_N$ & $\approx6.214\times 10^5 sec^{-1}$\tabularnewline
\hline
\end{tabular}

\caption{The set of local parameters used in a standard case microinstability
simulation based on the QAS3-C82 equilibrium. Note: $a_{N}$
is not the minor radius; it is discussed in App. A.}

\end{table}

Previously, FULL scans showed that the largest linear growth rate
occurs at  flux surface label $s=0.875$ (corresponding to a minor radius of $r/a\approx \sqrt{s} \approx 0.94$), for $\alpha=\pi/3$ and $\theta_{0}=0$. GS2
and FULL scans over $\alpha$ and $\theta_{0}$ (Figures 5 and 6)
adopted this $s$ value. The toroidal mode number, $n$, was fixed
at $25$ (thus, $k_{y}\rho_{i}=\frac{n}{a_{N}}\rho_{i}$
varied for each data point, because from App. A, $a_N=1/|\nabla \alpha|$ and $\rho_i \propto 1/B_a$ vary). These figures indicate good agreement
between the GS2 code and the FULL code. The maximum growth rate in
Figure 5 occurs for $\alpha=\pi/3$, and GS2 and FULL agree well around
this value. In Figure 6, GS2 and FULL again agree well around the
growth rate peak at $\theta_{0}=0$. 

In all further calculations presented in this paper, $s=0.875$, $\alpha=\pi/3$ and $\theta_{0}=0$, the location
of the maximum growth rate.
\begin{figure}
\includegraphics[scale=1.0]{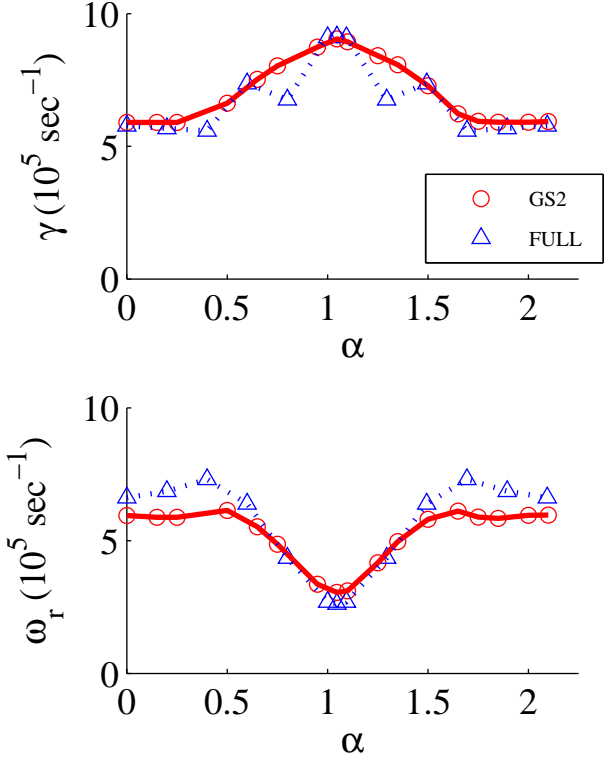}\caption{(color online) Variation of $\gamma$ and $\omega_{r}$ with $\alpha$ at constant
$s=0.875$ and $\theta_{0}=0$ with $\eta_{i}=\eta_{e}=3$ and $k_{y}\rho_{i}(\alpha=\frac{\pi}{3})=0.3983$.}

\end{figure}

\begin{figure}
\includegraphics[scale=1.0]{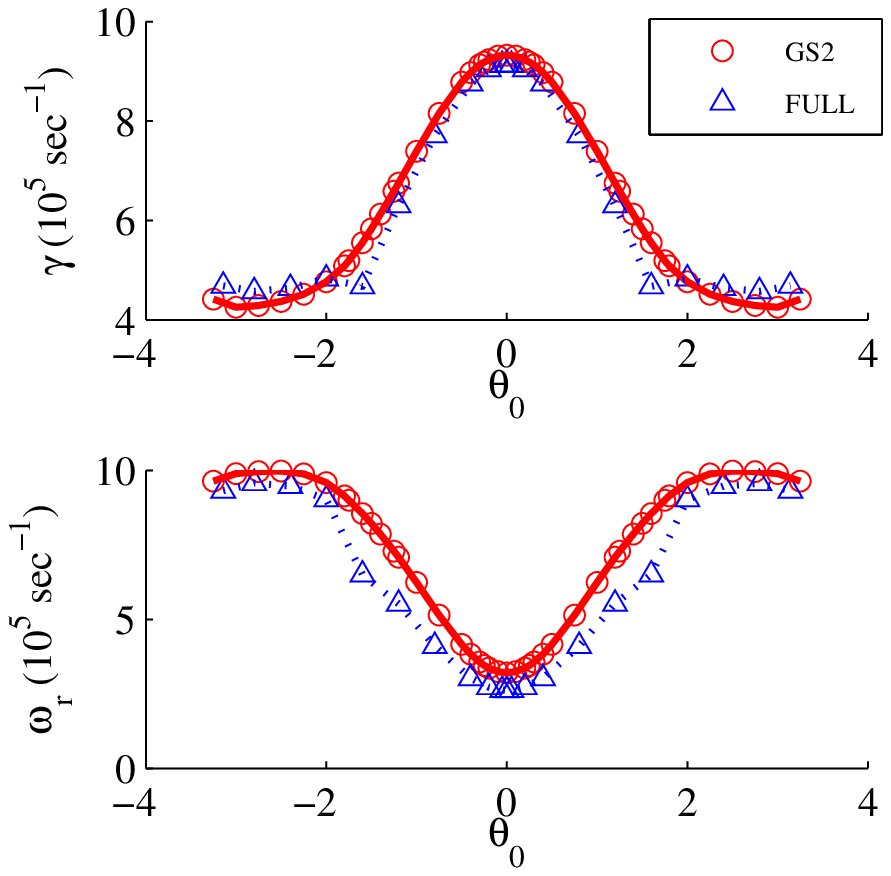}\caption{(color online) Variation of $\gamma$ and $\omega_{r}$ with $\theta_{0}$ at constant
$s=0.875$ and $\alpha=\pi/3$ with $\eta_{i}=\eta_{e}=3$ and $k_{y}\rho_{i}=0.3983$. }

\end{figure}

\begin{figure}
\includegraphics[scale=1.0]{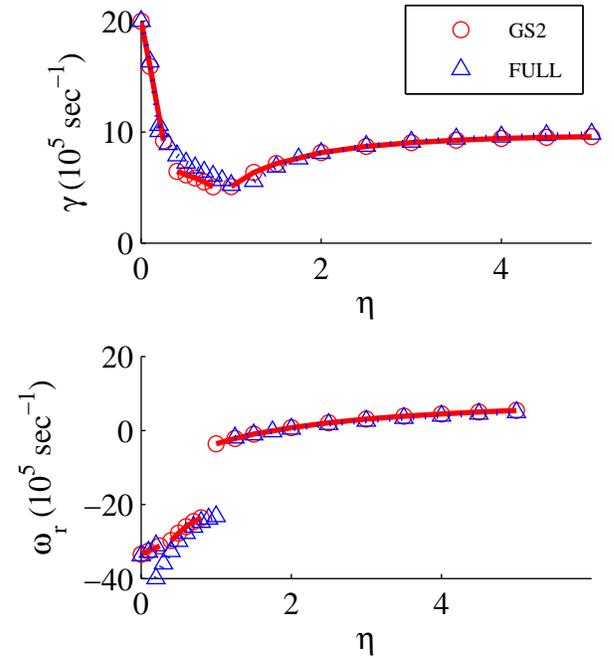}\caption{(color online) Variation of $\gamma$ and $\omega_{r}$ with $\eta_{i}=\eta_{e}$
with $k_{y}\rho_{i}=0.3983$.}

\end{figure}

We used GS2 to find the instability growth rate dependence on $\eta=L_{n}/L_{T}$
and compared it with FULL. The total pressure gradient was kept fixed
to maintain consistency with the MHD equilibrium. Both codes found
large growth rates at low $\eta$ (high density gradient) and high
$\eta$ (high temperature gradient) (Figure 7), and agree well, though
it can be seen in the frequencies that GS2 found a mode switch earlier
than FULL. This can happen since GS2 automatically finds the most unstable mode, whereas FULL usually finds the mode closest to the initial guess provided to the root finder. In fact, there are three distinct eigenmodes within these
regimes of $\eta$: at small $\eta$, even-symmetry TEM modes dominate;
at medium $\eta$, odd-symmetry TEM modes dominate; and at larger
values of $\eta$, an even-symmetry ITG-driven mode dominates\cite{rewoldt_drift_1999}
(Figure 8). This is typical of an equivalent axisymmetric configuration.\cite{rewoldt_toroidal_1990}

\begin{figure}
\includegraphics[scale=1.0]{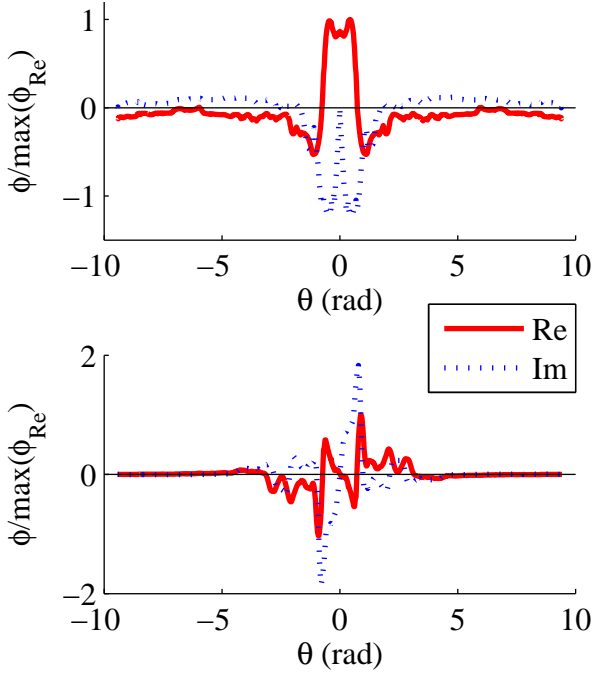}\caption{(color online) Variation of the normalized GS2 eigenfunctions of electrostatic, collisionless
toroidal drift modes along the field line  at $\eta=3$ (top figure) and at $\eta=0.5$
(bottom figure) with $k_{y}\rho_{i}=0.3983$.}

\end{figure}

Benchmarks with FULL for scans over $T_{i}/T_{e}$, shown in figure
9, were also successful. For this scan, $T_{e}$ was varied
while $T_{i}$ was kept constant at $1keV$. As $T_{i}/T_{e}$ increases, at this very large value of $R/L_{T_i}\approx 157$, the linear growth rate falls slowly due, most likely, to an enhancement of shielding by adiabatic electrons at large $\sqrt{T_i/T_e}$. This is a very well-known phenomenon
in tokamaks.

\begin{figure}
\includegraphics[scale=1.0]{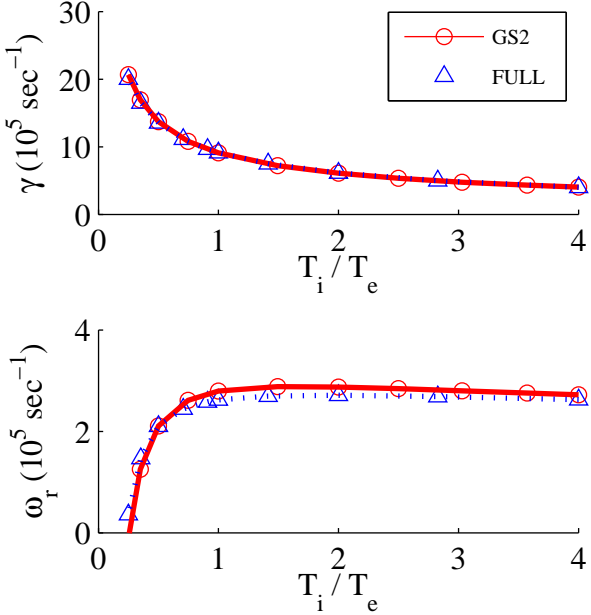}\caption{(color online) Variation of $\gamma$ and $\omega_{r}$ with $T_{i}/T_{e}$ with
$k_{y}\rho_{i}=0.3983$ and $\eta_{i}=\eta_{e}=3$.}

\end{figure}

Comparison scans over $k_{y}\rho_{i}$ for $\eta=0$ and
$\eta=3$ are shown in figure 10. For the $\eta=0$ curve, the dominating
eigenmodes are even in the ranges $0.1<k_{y}\rho_{i}<0.2$ and $0.6<k_{y}\rho_{i}<1.1$.
Overall, the results from the GS2 code and the FULL code agree well;
growth rates differ by at most $\sim10\%$ except at transitions
between modes.
\begin{figure}
\includegraphics[scale=1.0]{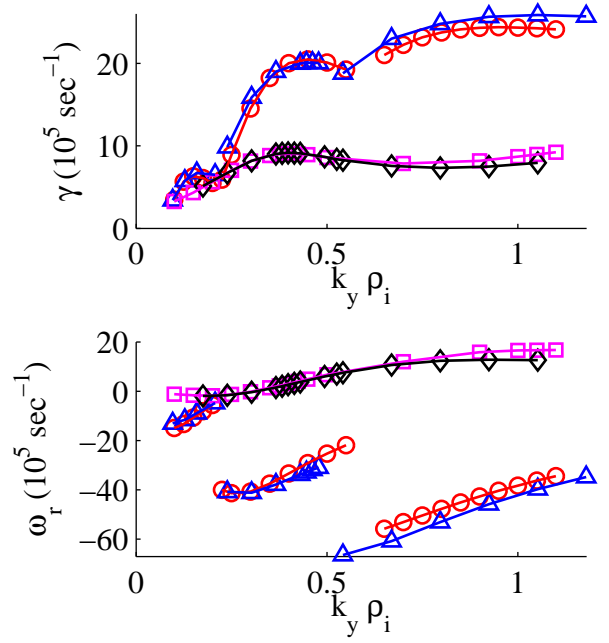}\caption{(color online) Variation of $\gamma$ and $\omega_{r}$ with $k_{y}\rho_{i}$. Circles: GS2, $\eta=0$; triangles: FULL, $\eta=0$; squares: GS2, $\eta=3$; diamonds: FULL, $\eta=3$.}

\end{figure}

We found high frequency, electron-temperature-gradient-driven (ETG) modes with GS2 at short wavelengths (Figure
11) in the extended $k_{y}\rho_{i}$ spectrum for the case of $\eta=3$.
This was not checked with FULL.

\begin{figure}
\includegraphics[scale=1.0]{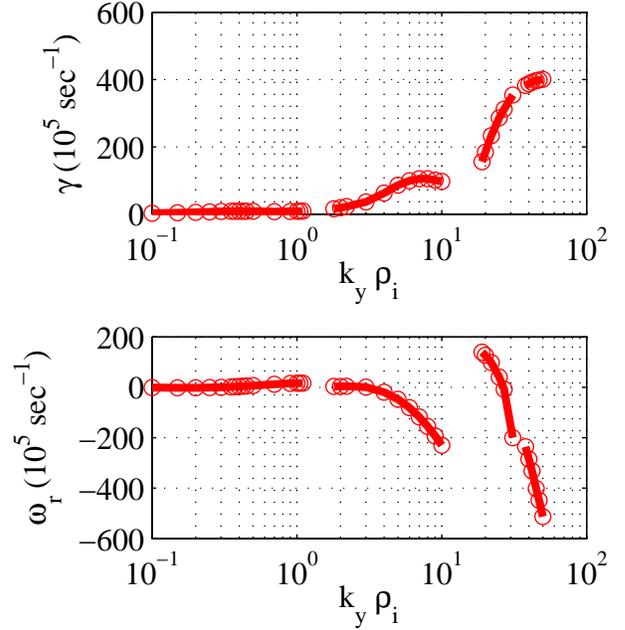}\caption{(color online) Extended variation from GS2 of $\gamma$ and $\omega_{r}$ with $k_{y}\rho_{i}$ and $\eta_{i}=\eta_{e}=3$.}

\end{figure}

\section{Critical Gradients for Linear Instability}

GS2 was also used to search for critical density gradients and temperature
gradients; i.e. to see whether gradients exist at which all drift
wave modes are stabilized. Note that for the next series of figures,
the normalizing length for the density and temperature gradient length
scales is defined as $a_{N}=(n/k_{\perp})(\theta=0)\sim0.352m$ (see App. A).

\begin{figure}
\includegraphics[scale=1.0]{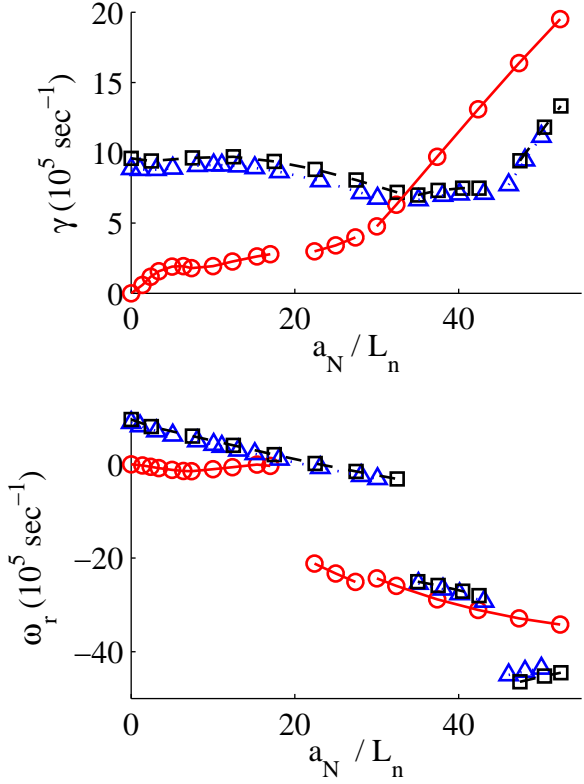}\caption{(color online) Variation of $\gamma$ and $\omega_{r}$ with $a_{N}/L_{n}$ with
$k_{y}\rho_{i}=0.3983$. Circles: $a_N/L_T=0.0$; triangles: $a_N/L_T=39.3$; squares: $a_N/L_T=44.9$.}

\end{figure}

Figure 12 shows a scan over the density gradient at various ion and
electron temperature gradients. The results are inconsistent with
the equilibrium pressure gradient, as the density gradient was increased
at constant temperature gradient. However, because the equilibrium
beta is so small ($\sim0.01\%$), the effect of the variation of the
pressure gradient is negligible. We see that there is no nonzero critical
density gradient threshold, even in the absence of temperature gradients.
There are switches in eigenmode symmetry from even to odd as $a_{N}/L_{n}$
increases, or all $a_{N}/L_{T}$ values.

\begin{figure}

\includegraphics[scale=1.0]{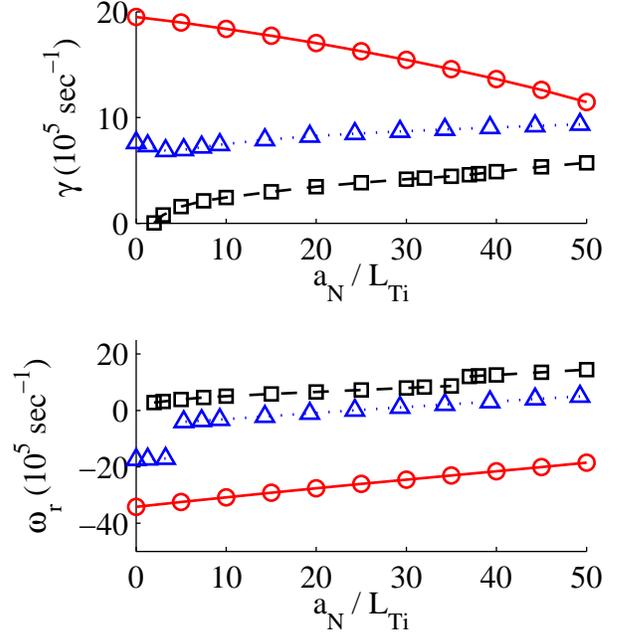}\caption{(color online) Variation from GS2 of $\gamma$ and $\omega_{r}$ with $a_{N}/L_{Ti}$ with
$k_{y}\rho_{i}=0.3983$. Circles: $a_N/L_n=52.4, a_N/L_{Te}=0.0$; triangles: $a_N/L_n=13.1, a_N/L_{Te}=39.3$; squares: $a_N/L_n=0.0, a_N/L_{Te}=0.0$.}

\end{figure}

\begin{figure}

\includegraphics[scale=1.0]{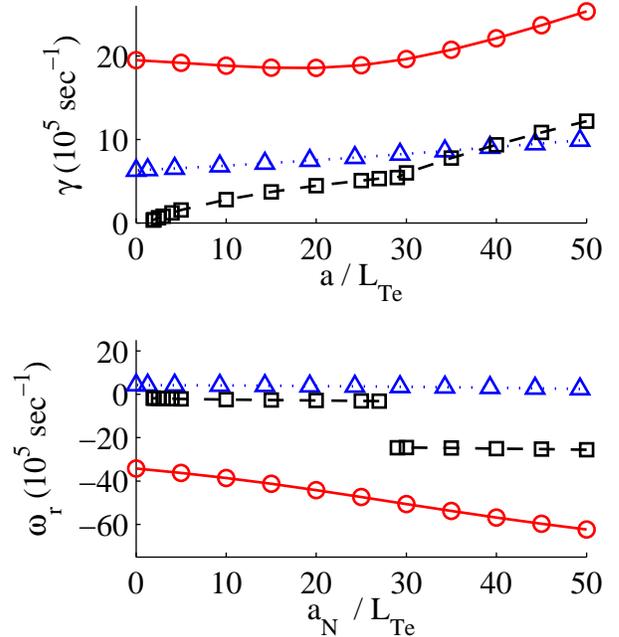}\caption{(color online) Variation from GS2 of $\gamma$ and $\omega_{r}$ with $a_{N}/L_{Te}$ for
the case of Fig. 2 with $k_{y}\rho_{i}=0.3983$. Circles: $a_N/L_n=52.4, a_N/L_{Ti}=0.0$; triangles: $a_N/L_n=13.1, a_N/L_{Ti}=39.3$; squares: $a_N/L_n=0.0, a_N/L_{Ti}=0.0$.}

\end{figure}

However, a critical ion temperature gradient for an ITG-driven mode
was found at $a_{N}/L_{Ti}\sim2$ (or $\frac{R}{L_{T_{i}}}=\frac{R}{a_{N}}\frac{a_{N}}{L_{T_{i}}}\approx4\frac{a_{N}}{L_{T_{i}}}=8$)
in the absence of all other gradients (Figure 13). Likewise,
a critical electron temperature gradient for a TEM-driven mode was
found at $\frac{a_N}{L_{T_{e}}}\sim2$ in the absence of all other gradients
(Figure 14).

\section{Conclusions}

The nonlinear gyrokinetic code GS2 has been extended to treat non-axisymmetric
stellarator geometry. Geometric quantities required for the gyrokinetic
simulations are calculated from a VMEC-generated equilibrium using
the VVBAL code and are further described in App. A. 

Linear, collisionless, electrostatic simulations of the quasi-axisymmetric,
three-field period NCSX stellarator design QAS3-C82 have been successfully
benchmarked with the eigenvalue code FULL for scans over a range of
parameters including $\eta$, $k_{y}\rho_{i}$, $T_{i}/T_{e}$,
$\alpha$, and $\theta_{0}$. Quantitatively, the linear stability
calculations of GS2 and FULL agree to within about $10\%$ of the
mean, except at transitions between modes. Further results using only GS2 included short wavelength
modes, odd parity, faster growing modes, and the effect of individual
density and temperature gradients.

Future work will include the exploration of the effects of collisionality
and electromagnetic dynamics, investigation of finite beta equilibria,
and, most significantly, the effects of nonlinear dynamics. A benchmark
of stellarator studies is underway between GS2 and the continuum gyrokinetic
code GENE\cite{jenko_electron_2000} for NCSX, as well as stellarators
W7-AS and W7-X. 

GIST\cite{xanthopoulos_geometry_2009} is now capable of creating
GS2 geometry data files and will be used in the future, along with the new GS2 grid generator. 

\begin{acknowledgements}
The authors would like to thank Dr. L.-P. Ku and Dr. W. A. Cooper
for providing equilibrium results from the VMEC and TERPSICHORE codes
as well as the resulting parameters from the VVBAL code.  Also, thank you to M. A. Barnes for useful discussion.

This work was supported by the U.S. Department of Energy through the 
SciDAC Center for the Study of Plasma Microturbulence and the Princeton 
Plasma Physics Laboratory by DOE Contract No. DE-AC02-09CH11466, and by 
a DOE Fusion Energy Sciences Fellowship.

\end{acknowledgements}

\appendix
\section{Geometry Details\label{Appendix A} }
In order to make the simulation grid for these GS2 stellarator runs,
VMEC creates the 3-D MHD equilibrium, TERPSICHORE transforms it into
Boozer coordinates, and VVBAL calculates necessary geometric coefficients
along a specified field line. Then, GS2's grid generator, Rungridgen,
creates the final grid for use in the microinstability calculations. (A new grid generator is in production, which will be used for further GS2 stellarator calculations.)
The normalizations of geometric quantities change between these codes,
and knowing them in detail is required for benchmarks between gyrokinetic
codes. We define the normalizing length, $a_N$, in App. A.2.

In GS2, the field-aligned coordinate system is $(\rho,\alpha,\theta)$.  $\theta$ is the poloidal angle and distance along the field line. The magnetic field takes the form $\mathbf{B}=\nabla\alpha\times\nabla\Psi$, where $\alpha=\zeta-q\theta$ is the field line label. The radial coordinate, $\rho$, can differ between codes, and we define it in App. A.1. More details of general geometry for GS2 are documented in App. A of Ref. \onlinecite{barnes_trinity:_2009}.

\subsection{Radial coordinate, $\rho$ \label{sub:Radial-coordinate}}

VMEC and TERPSICHORE use the normalized toroidal flux surface label
$s=\Phi/\Phi_{edge}\sim\langle(r/a)^{2}\rangle$ as the radial coordinate,
$\rho$. In the customized version of VVBAL used here, the radial coordinate is transformed
to $\rho=\Psi_{N}=\Psi/(a_{N}^{2}B_{a})$, where $\Psi_{N}$ is the
normalized poloidal flux. 

Because Rungridgen uses VVBAL output without modification, here $d\rho/d\psi_{N}\equiv1$.
(In Ref. \onlinecite{barnes_trinity:_2009}, the
definition of the geometry coefficients include the variable $d\rho/d\psi_{N}$,
which can be used to choose the radial coordinate.)

\subsection{Normalizing Quantities, $B_a$ and $a_N$ \label{sub:Norm-Quantities}}

The normalizing magnetic field is $B_{a}=\langle B\rangle$,  where $\langle B\rangle$ is a theta-average, not weighted to be a flux-surface average (Ref. \onlinecite{barnes_trinity:_2009} chooses $B_a$ differently).  

The normalizing length is  $a_{N}$, given for these calculations by VVBAL as
\begin{equation}
a_{N}=\frac{n}{\sqrt{|\mathbf{k_{\perp}}|^{2}(\theta=0, \theta_0=0)}}=\frac{1}{|\nabla \alpha|}.
\end{equation}

GS2 treats perturbed quantities as $A=\hat{A}(\theta)exp(iS)$,
where  $\mathbf{k_{\perp}}=\nabla S=n\nabla(\alpha+q\theta_{0})=n\nabla[\zeta-q(\theta-\theta_0)]$; $n$ is the toroidal mode number. (In non-axisymmetric devices, $n$ is not a conserved quantum number, because toroidal variations in the equilibrium give coupling between $n$ modes. However, in the small-$\rho*$, high-$n$ limit, this coupling is weak, and $n$ can just be considered a coefficient to select a particular value of $k_\perp$.) 

In the notation of Eqn. A.11 of App. A in Ref. \onlinecite{barnes_trinity:_2009},
\begin{equation}
|\mathbf{k_{\perp}}|^{2}=|\nabla S|^{2}=k_{\theta}^{2}\left|g_{1}+2\theta_{0}g_{2}+\theta_{0}^{2}g_{3}\right|
\end{equation}

where $g_1$, $g_2$, and $g_3$ are coefficients in the geometry file written by VVBAL and read by GS2. Also, $k_\theta=k_y=n/a_N$. (The GS2 variable \texttt{aky} is defined as $k_y\rho_i$, with $\rho_i\propto 1/B_a$.)

In the notation of Eqn. 7 of Ref. \onlinecite{xanthopoulos_geometry_2009},

\begin{equation}
|\mathbf{k_{\perp}}|^{2}
=n^2\frac{\sqrt{g}B^{2}}{\Psi'^{2}(s)}[C_{p}+C_{s}(\theta-\theta_0)+C_{q}(\theta-\theta_0)^{2}], 
\end{equation}
where $\sqrt{g}$ is the Jacobian,  $C_{p}$, $C_{s}$, and $C_{q}$
are defined in section II of  Ref. \onlinecite{xanthopoulos_geometry_2009}.

So, VVBAL writes:
\begin{equation}
g_{1}=a_{N}^{2}\frac{\sqrt{g}B^{2}}{\Psi'^{2}(s)}[C_{p}+C_{s}\theta+C_{q}\theta^{2}]
\end{equation}

\begin{equation}
g_{2}=-a_{N}^{2}\frac{\sqrt{g}B^{2}}{\Psi'^{2}(s)}\left[C_{q}\theta+\frac{C_{s}}{2}\right]
\end{equation}

\begin{equation}
g_{3}=a_{N}^{2}\frac{\sqrt{g}B^{2}}{\Psi'^{2}(s)}C_{q}
\end{equation}

\end{document}